\documentclass[aps,pra,reprint,superscriptaddress,showpacs]{revtex4-1} 
\usepackage{graphicx}  
\usepackage{dcolumn}   
\usepackage{bm}        
\usepackage{verbatim}   
\usepackage[version=3]{mhchem} 
\usepackage{amsmath,amssymb}
\usepackage{textcomp} 
\usepackage{enumitem}
\usepackage{color}
\usepackage{psfrag}
\usepackage[breaklinks,colorlinks,citecolor=blue,linkcolor=blue,urlcolor=blue]{hyperref}

\begin{document}

\title{Simultaneous Zeeman deceleration of polyatomic free radical with lithium atoms }

\author{Yang Liu}
\email[]{liuyang59@mail.sysu.edu.cn}
\affiliation{School of Physics and Astronomy, Sun Yat-Sen University, Zhuhai, 519082, China}

\author{Luo Le}
\email[]{luole5@mail.sysu.edu.cn}
\affiliation{School of Physics and Astronomy, Sun Yat-Sen University, Zhuhai, 519082, China}

\date{\today}

\begin{abstract}

Chemistry in the ultracold regime enables fully quantum-controlled interactions between atoms and molecules, leading to the discovery of the hidden mechanisms in chemical reactions which are usually curtained by thermal averaging in the high temperature. Recently a couple of diatomic molecules have been cooled to ultracold regime based on laser cooling techniques, but the chemistry associated with these simple molecules is highly limited. In comparison, free radicals play a major role in many important chemical reactions, but yet to be cooled to submillikelvin temperature. Here we propose a novel method of decelerating \ce{CH3}, the simplest polyatomic free radical, with lithium atoms simultaneously by travelling wave magnetic decelerator. This scheme paves the way towards co-trapping \ce{CH3} and lithium, so that sympathetical cooling can be used to preparing ultracold free radical sample.

\end{abstract}

\maketitle

\vspace{-1mm}

\noindent\textit{\textbf{Introduction}}\quad The field of physical chemistry or chemical physics have seen astonishing strides towards the creation of cold and ultracold molecules in electronic and rovibrational ground state in recent two decades. Such research have fostered a wealth of interdisciplinary explorations, such as many-body quantum physics and chemistry \cite{baranov2008theoretical,eisert2015quantum}, quantum computation \cite{demille2002quantum,rabl2006hybrid}, quantum simulation \cite{micheli2006toolbox,gorshkov2011tunable,yan2013observation}, cold and ultracold chemistry\cite{balakrishnan2001chemistry,krems2008cold,bell2009ultracold,ospelkaus2010quantum,stuhl2014cold,dulieu2017cold}, precision measurement \cite{hudson2006cold,zelevinsky2008precision,chin2009ultracold,kobayashi2019measurement,baron2014order,demille2017probing,andreev2018improved} . Although a couple of diatomic molecules have already been successfully cooled to ultracold regime, these simple molecules are lack of potential to explore rich chemistry in a general sense. In comparison, free radicals involving in many crucial chemical reactions, but yet to be cooled to submillikelvin temperature. For example, methyl radical \ce{CH3}, the simplest organic polyatomic radical, is one of the most important and fundamental intermediates in hydrocarbon chemistry. It plays a key role in various reactions including combustion, atmospheric and interstellar chemistry. Creating ultracold \ce{CH_{3}} would help to understand the quantum mechanisms related to many elementary reactions. For example, at very low temperature, two types of reactions could happen, one is barrierless reaction \ce{CH_{3} + OH -> CH_{2}O + H_{2}}, and the other is tunnelling process, \ce{CH_{3} + H_{2} -> CH4 + H} \cite{momose1998tunneling,hoshina2004tunneling}. However their reaction rate and branching ratio are still ambiguous \cite{jasper2007kinetics}. Understanding these reactions would give a thrust to the advancement of cold chemistry.

\ce{CH3} molecule has an unpaired electron, and has a linear Zeeman shift in strong magnetic fields because its spin is decoupled from the molecular axis and preferentially oriented relative to the external field. An efficient cooling method for \ce{CH3} is translational deceleration and trapping by a well-designed time-varying magnetic \cite{momose2013manipulation,liu2015one}, which can be followed by a second-stage cooling \cite{stuhl2012evaporative}. However, one of the fundamental requirement for the second-stage cooling to proceed is a much larger molecular density than what can be achieved by the usual decelerator. In a recent work, \ce{CH3} molecules with a translational temperature of 200 mK are obtained and are trapped for more than 1 s, but the estimated density is on the order of $5.0\times 10^{7} cm^{-3}$ \cite{liu2017magnetic}. This density is not large enough to ensure further cooling of \ce{CH3} molecules to ultracold regime even if assuming favourable ratio of elastic to inelastic collision cross section between \ce{CH3} molecules. \\[3pt]

\noindent\textit{\textbf{Co-Deceleration Scheme}}\quad In order to improve the \ce{CH3} density, and also inspired by the recent studies on the collision of the \ce{O2-Li} mixture \cite{akerman2017trapping}, here we propose to simultaneously decelerate \ce{CH3} molecule with Li atom using the moving-magnetic trap decelerator \cite{lavert2011moving,lavert2011stopping}. This method has two unique advantages. It genuinely has a larger deceleration efficiency over conventional Zeeman deceleration due to larger phase space acceptance for small final velocity, and inherently smoothing deceleration with true three-dimensional trapping potentials, meanwhile it offers simultaneous deceleration of \ce{CH3} and Li since this deceleration scheme does not rely on the ratio of mass to magnetic dipole moment. The simple electronic configuration and light mass of Li atom make study of atom-molecule collisions much less complicated. Thus the decelerated mixture can be used as an ideal test system for sympathetic cooling of \ce{CH3} molecules.

\begin{figure}
\begin{center}
\includegraphics[width=1.0\linewidth]{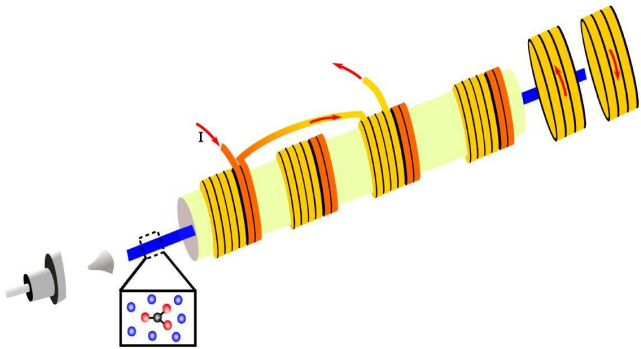}
\caption{The proposed experimental scheme. The mixture of \ce{CH3} molecules and Li atoms ejected from a supersonic valve, pass a skimmer, and enters a vacuum tube wrapped with an array of coils, which constitute the moving Zeeman decelerator. The red arrows represent the current flow, and the current in yellow coils flow in opposite direction to the current in orange coils. The \ce{CH3} molecules and Li atoms are decelerated simultaneously in the decelerator. After the deceleration, they will be co-trapped with a magnetic trap consisting a pair of anti-Helmholtz coils. }
\label{scheme}
\end{center}
\end{figure}

The scheme is shown in Figure ~\ref{scheme}, which is based on the moving magnetic trapping technique as demonstrated in \cite{lavert2011moving}. The supersonic beam of \ce{CH3} molecules and Li atoms can be created from a pulsed valve of Even-Lavie type, whose temperature can be continuously adjusted through a liquid nitrogen flowing jacket \cite{liu2017magnetic}, therefore enabling smoothly velocity tuning of the beam. Both \ce{CH3} molecules and Li atoms would be seeded in either krypton or xenon gas to achieve a small mean velocity. Using a mixture of $15\%$ \ce{CH4} seeded in Kr, the supersonic beam of \ce{CH3} can be created by DC electric discharge, and have a mean velocity of 330 m/s with a standard deviation of $18\%$ using seed gas of krypton \cite{momose2013manipulation,liu2017magnetic}. Lithium atoms can be entrained into the beam post-nozzle by picking up laser-ablated lithium \cite{akerman2017trapping} or effusive lithium atoms from a heated oven \cite{jerkins2010efficient,melin2019observation}. As the \ce{CH3-Kr} mixture expands over the supersonic expansion, the lithium mixes and assumes the temperature and spatial profile of the \ce{CH3-Kr} mixture. After passing through a 5 mm skimmer downstream, the beam enters the decelerator.

In the molecular beam experiments \cite{momose2013manipulation,liu2017magnetic}, the typical dimensions of the pulsed methyl radical beam is quite long in beam propagation direction and roughly expand over $4mm\times 4mm\times 3cm$. In order to create a trapping region which can load the central part of the beam, we model our trap from two coils with a inner diameter of 4mm and a center-to-center distance of 10mm, which can be wrapped around a thin-walled glass tube used as the deceleration channel of the mixture. The coil geometry is designed as 16 turns ($4\times 4$) of a 25 gauge (AWG) Cu wire for the first coil whereas the second has 8 ($4\times 2$). The decelerator consists of 198 overlapping quadrupole traps overall thus extending over  997.7 mm.

In Figure ~\ref{scheme} the red arrows represent the current flow, and the current in yellow coils flow in opposite direction to the current in orange coils, thus creating a quadrupole magnetic trap in this anti-Helmholtz configuration. Such pair of coils are arranged in series along the atomic/molecular beam axis, so the quadrupole magnetic traps created by neighbouring pairs are spatially overlapping with each other. Considering two neighbouring pairs, current pulse of first pair has a half sine shape profile with a pulse width of $\tau/2$ and the delay time between these two successive pulses is $\tau/4$, which means the second current pulse is send through the second pair when the current flowing through the first pair reaches its maximum. Afterwards the current decreases in first pair and increases in the second pair. Thus their current pulse sequences are temporally overlapping, creating a three dimensional moving magnetic trap.

\begin{figure}
\begin{center}
\includegraphics[width=1.0\linewidth]{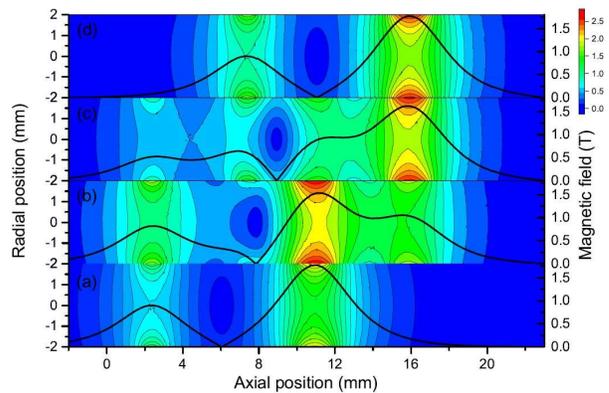}
\caption{Cross-section of spatial magnetic field distribution at $t=\frac{\pi}{2}$, $t=\frac{2\pi}{3}$, $t=\frac{5\pi}{6}$, and $t=\pi$ during the moving of the magnetic trap from bottom to top panel. Black solid lines and contour plots are the magnetic field distribution along axial direction and in a two-dimensional cut using finite element simulation, respectively. }
\label{field}
\end{center}
\end{figure}

We adapt sinusoidal current pulses as \cite{lavert2011moving}. They can be described using the form of
$I_{n}= I_{0}sin[\omega \cdot t - \frac{n\pi}{2}]$, where subscript n represents the n-th pair of anti-Helmholtz coils, and modulating frequency $\omega$ sets how fast the trap moves in time. Figure ~\ref{field} illustrates the resulting spatial magnetic field distribution at $t=\frac{\pi}{2}$, $t=\frac{2\pi}{3}$, $t=\frac{5\pi}{6}$, and $t=\pi$ during the moving of the magnetic trap from first pair to second pair, where $t=0$ corresponds to the time of activation for the first pair.

The beam will move at constant speed with fixed $\omega$, while the deceleration is accomplished by chirping $\omega$. The deceleration process depends on the frequency chirping rate $\frac{d\omega}{dt}$, which in turn determines current pulse timings and consequent time-varying magnetic fields. The timing of pulse sequence for each pair of anti-Helmholtz coils is calculated according to the initial and final velocity of the particle before and after the decelerator. Assuming a linear frequency chirp of the current pulse, the time-dependent phase of n-th pair can be approximated as $\Phi (t)= \int{\omega (t)dt}=\omega_{0}t+\pi a t^{2}=\frac{2\pi}{L} [v_{0}t-\frac{v_{i}^2-v_{f}^2}{4n}t^2]$, where $a$ is the deceleration value, $L$ is the distance between neighbouring pairs, $v_0$ is the velocity reaching the n-th pair, $v_i$ and $v_f$ corresponds to the velocity of the beam before and after the whole decelerator, respectively. Therefore, the full chirped current profile can be given as $I_{n}= I_{0}sin[\Phi - \frac{n\pi}{2}]$.\\[3pt]

\begin{figure}
\begin{center}
\includegraphics[width=1.0\linewidth]{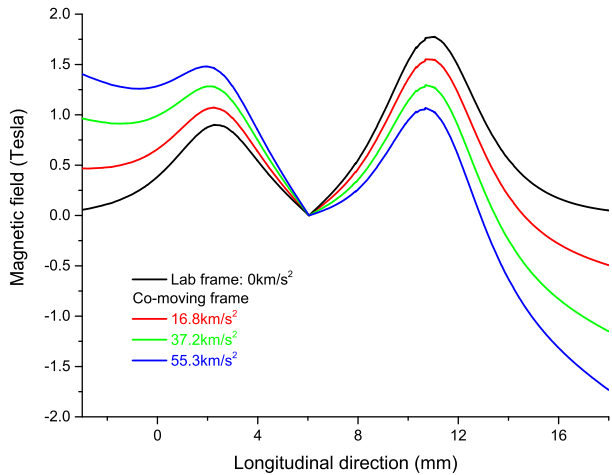}
\caption{Modified potential along the longitudinal direction in the lab frame(black), in the moving frame with decelerations of $a=16.8 km/s^2$(red), $a=37.2 km/s^2$(green), $a=55.3 km/s^2$(blue), respectively.}
\label{potential}
\end{center}
\end{figure}

\noindent\textit{\textbf{Deceleration Dynamics}}\quad In this section, we show the simulation of the one-dimensional simultaneous deceleration of \ce{CH3} and Li. The deceleration process in 1D can be approximated by a fictitious time-independent conservative
force to a first approximation. It generates a scalar potential:
\begin{equation} \label{force}
F_{fic}=-\frac{\partial{(W^{\prime}+W_{0})}}{\partial z}=-\frac{\partial{\mu_{eff}(B^{\prime}+B_{0})}}{\partial z}=ma
\end{equation}
which tilts the magnetic field potential in the decelerating frame, such that it lowers the front barrier and increases the back barrier, as seen from Figure ~\ref{potential}. The dynamics of the particle inside the trap during deceleration can be approximated by
\begin{equation} \label{dynamics}
m\frac{d^{2}z}{dt^{2}}=\frac{2\pi}{L}W_{max}[cos(kz)+\epsilon]
\end{equation}
where $W_{max}$ is the trap depth, $k=\frac{2\pi}{L}$, and $\epsilon$ is a factor proportional to the deceleration value.

In the following, we first show how the magnetic potential is modified according to different acceleration value, based on which we give the Monte-Carlo simulation of the deceleration dynamic process, resulting in the simulated time-of-flight spectrum. Then we obtain the dependence of deceleration efficiency on the final target velocity.

We use a current of 500 A which provides a field magnitude of 1.8 T at the front barrier and 0.9 T magnitude at the back. Figure ~\ref{potential} shows the modified longitudinal field $B^{\prime}$ due to deceleration for several deceleration values. In the case of \ce{CH3} deceleration with 55.3 $km/s^2$ the fictitious force adds a negative 48 T/m tilt to the 415 T/m of the initial magnetic potential, reducing the height of the front barrier to 1.28 T, which is equivalent to a trap depth of 860 mK.

With these calculation, we model the motion of the particles inside the decelerator. We have calculated their trajectories using Monte-Carlo simulations without any free parameters. In the simulations a million particles are assumed to be Gaussian distributed at the beam origin position where \ce{CH3} are produced by the discharge for both \ce{CH3} and Li. Based on previous experiments \cite{momose2013manipulation,liu2017magnetic}, nearly all the produced radicals are populated in the lowest rotational state $|N=0,K=0\rangle$ and $|N=1,|K|=1\rangle$, which belong to ortho and para type according to nuclear spin statistics, respectively. Assuming the same production efficiency of $7.36\%$ as NH from \ce{NH3}\cite{luria2009dielectric}, together with the seeding ratio of $15\%$, we have $1.1\%$ of \ce{CH3} in the beam. If we use a lower bound $5\%$ \cite{jerkins2010efficient} as the entraining efficiency of lithium atoms into the supersonic beam, then the concentration ratio of lithium to \ce{CH3} is about 5:1.

\begin{figure}
\begin{center}
\includegraphics[width=1.0\linewidth]{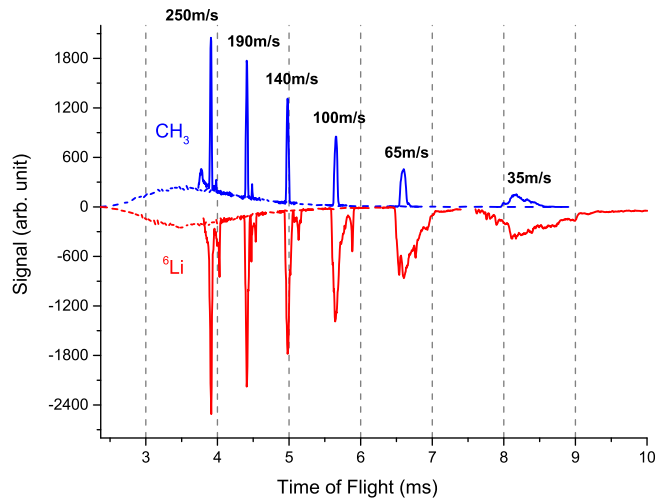}
\caption{Deceleration of \ce{CH3} and Li. Time-of-flight traces of \ce{CH3}(blue) and Li (red) for various final velocities, from free flight with mean velocity of 340m/s down to 35m/s.}
\label{velocity}
\end{center}
\end{figure}

In current simulation, we take the concentration ratio as 1:1 for convenience. Assuming rotational population of the \ce{CH3} radical on the lowest rotational states of each nuclear spin isomers, and population of lithium atom on the lowest six hyperfine states follow Boltzmann distribution, we have performed simulations for six final velocities, 250 m/s, 190 m/s, 140 m/s, 100 m/s, 65 m/s, and 35 m/s. The resulting time-of-flight spectrum from one-dimensional simulation are presented in Figure ~\ref{velocity}, which consists of seven time-of-flight traces including the free flight (dashed line) and decelerated packet at various final velocities for both \ce{CH3}(blue line) and Li(red inverted line). We notice \ce{CH3} molecule and Li atom arrive at roughly the same time, which clearly indicate they have been simultaneously decelerated. The observed small difference is due to a longer deceleration time needed to decelerate to the final velocity as pointed out previously by \cite{lavert2011moving}.

We also obtain the relative number of decelerated \ce{CH3} molecules and Li atoms as a function of the final velocity from the simulation, which is shown in left panel of Figure ~\ref{particle}. Both curves were normalized to the final velocity of 255m/s. For both species the relative number show a monotonic dependence on the final velocity, the smaller the final velocity, the smaller the relative number of decelerated particles. Another feature is larger deceleration efficiency for lithium atoms than for \ce{CH3} radicals due to larger magnetic moment to mass ratio for lithium atom. A relative sharp decrease of deceleration efficiency when final velocity is lower than 65 m/s is because of the dependence of the effective magnetic potential on the deceleration value. The effective trapping potentials in the moving frame of reference with a final velocity of 100m/s for both species are plotted in right panel of Fig. 4. The trap depth of \ce{CH3} is more than 0.3 larger is again due to smaller mass to magnetic moment ratio.\\[3pt]

\begin{figure}
\begin{center}
\includegraphics[width=1.0\linewidth]{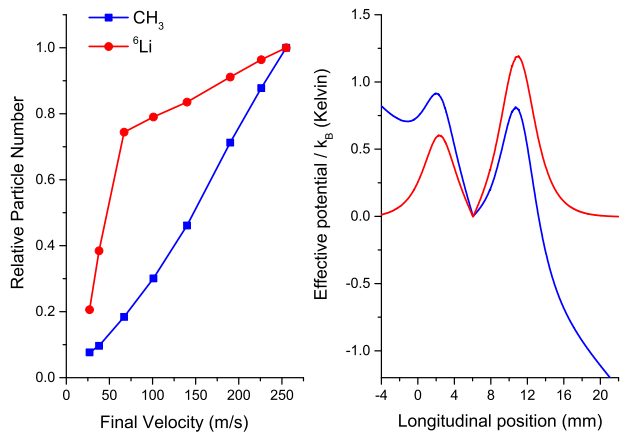}
\caption{Relative number of \ce{CH3} molecules (blue) and Li atoms (red) as a function of the final velocity. Both curves are normalized to the final velocity of 255m/s. (right) The effective trapping potentials for \ce{CH3} and Li in the moving frame of reference with final velocity of 100m/s.}
\label{particle}
\end{center}
\end{figure}

\noindent\textit{\textbf{\ce{CH3}-Li Collision Properties}}\quad  Study collisions between \ce{CH3} and Li would give an important hint for sympathetic cooling of \ce{CH3} molecules by Li atoms. T. V. Tscherbul et al. theoretically studied cold collisions of \ce{CH3} molecules and $^{3}$He using both unmodified and strongly anisotropic interaction potentials for He-\ce{CH2} \cite{tscherbul2011sympathetic}, which gives the ratio of the rate constant for elastic scattering and spin relaxation $9.8\times 10^{12}$ and $2.8\times 10^{7}$ at $T=0.5K$ and $B=0.1T$, indicating \ce{CH3} an promising candidate for sympathetic cooling experiments using cold $^{3}$He gas. Compared with $^{3}$He, lithium atom can be easily laser cooled, and is also an excellent coolant atoms according to recent theoretical calculations \cite{tscherbul2011ultracold,wallis2011prospects,morita2017cold}. In a magnetic trap, Timur V. Tscherbul et al. have shown the inelastic cross sections for interspecies collisions between $^{2}\Sigma$ molecular radicals and alkali-metal atoms are strongly suppressed due to the weakness of the spin-rotation interaction in $^{2}\Sigma$ molecules \cite{tscherbul2011ultracold}, and the spin-relaxation collisions would probably be suppressed between spin-stretched \ce{CH3} and Li, thus sympathetic cooling of methyl radical with laser-cooled lithium atoms is likely to be successful.

Here we use quantum diffractive scattering \cite{fagnan2009observation} to model the scattering between co-trapped \ce{CH3} and Li. Assuming the interaction between \ce{CH3} and Li is dominated by long-range van der Waals force, then the interaction potential can be model by an ideal Lennard-Jones potential, $V(r) = - \frac{C_{6}} {r^6}$ , where the value $C_6$ can be approximated by $C_{6} = \frac{3}{2}  \frac{I_{\ce{CH3}} I_{\ce{Li}}} {I_{\ce{CH3}}+I_{\ce{Li}}} \cdot  \alpha_{\ce{CH3}} \cdot  \alpha_{\ce{Li}}$ using London dispersion force. Here, $I_{\ce{CH3}}$ and $I_{\ce{Li}}$ are ionization energy of \ce{CH3} and \ce{Li}, respectively. $\alpha_{\ce{CH3}}$ and $\alpha_{\ce{Li}}$ is the polarizability of \ce{CH3} and Li, respectively. The scattering wavefunction and scattering amplitude can be expanded in terms of the Legendre polynomials $$\psi_{k} (r,\theta) = \sum^{\infty}_{l=0}  R_{l} (k,r)\cdot P_{l} (cos\theta)$$ and
\begin{eqnarray*}
f(k,\theta) & = & \sum^{\infty}_{l=0} {f_{l} (k) P_{l} (cos\theta)} \\
& = &  \sum^{\infty}_{l=0} {\frac{2l+1}{k} \cdot e^{i\delta_{l}}\cdot sin\delta_{l}\cdot P_{l}  (cos\delta)} ,
\end{eqnarray*}
respectively, where k is collision wave vector and $\delta_l$ is the phase shift of the $l$th partial wave.

The determination of the scattering amplitude and resultant collision cross section $$\sigma(k)=\int_{0}^{\pi}{2\pi|f(k,\theta)|^2 sin\theta d\theta}$$ requires finding the partial wave phase shifts, which can be obtained by numerical integration of the radial Schrodinger equation $$(\nabla^2-\frac{2\mu}{\hbar^2} U(r)+k^2 )\psi(r)=0.$$ The solution to the radial equation for each partial wave $l$ is independently computed using the logarithmic-derivative method.

In order to study the diffractive scattering between \ce{CH3} and Li, and the prospects for further sympathetic cooling, we assume lithium atoms in the trap are laser cooled to a temperature of $500\mu K$ following magnetic trapping. Fig.1 shows the theoretically computed total cross section for the \ce{CH3}-Li collisions, which is averaged over a normal velocity distribution at 200mK. The inset is the partial cross section $$\sigma_{l} (k)=\frac{4\pi(2l+1)}{k^2} \cdot sin^2 \delta_{l}$$ as a function of the partial wave value $l$, which exhibit a universal shape between $L=98\hbar$ and $L=110\hbar$ and core dependent oscillations below. Fig. 2 is a plot of the velocity-averaged loss rate constant versus collision energy for the \ce{CH3}-Li collisions.

\begin{figure}
\begin{center}
\includegraphics[width=1.0\linewidth]{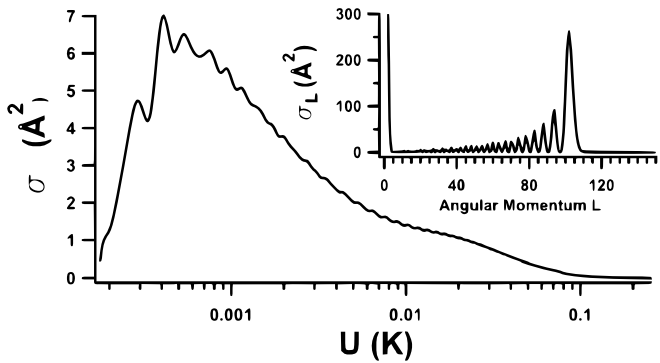}
\caption{Theoretically computed total cross section for the \ce{CH3}-Li collisions. The inset shows the partial cross section $\sigma_l$ as a function of the partial wave value $l$.}
\label{particle}
\end{center}
\end{figure}

\begin{figure}
\begin{center}
\includegraphics[width=1.0\linewidth]{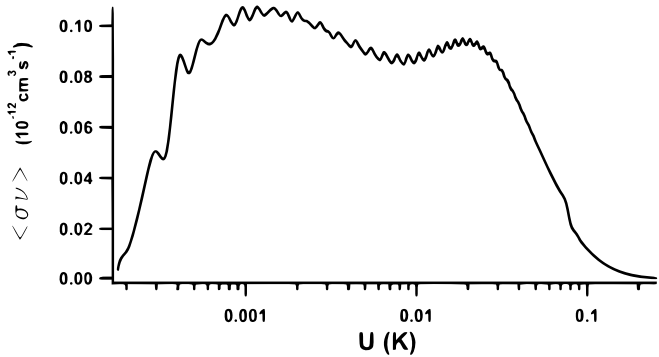}
\caption{The theoretically computed loss rate constant $\langle \sigma \cdot v\rangle_{Li,\ce{CH3}}$ versus collisional energy for laser-cooled Li atoms and magnetic trapped \ce{CH3} radicals.}
\label{particle}
\end{center}
\end{figure}

More accurate calculation of scattering cross sections would require not only highly accurate potential energy surfaces constructed by high-level \textit{ab initio} electronic structure calculations such as coupled cluster method with single, double, and perturbative triple excitations[CCSD(T)], but also multi-channel scattering calculation  \cite{wallis2011prospects}, where scattering cross sections between levels $i$ and $f$ are given by $$\sigma_{i\rightarrow f} = \frac{\pi} {k_{i}^2}  \sum_{l,m_{l}}\sum_{l^{'},m_{l}^{'}} |\delta_{i,f} \delta_{l,l^{'} } \delta_{m_{l},m_{l}^{'} } - S_{ilm_{l}; fl^{'},m_{l}^{'} }|^2.$$

\noindent\textit{\textbf{Conclusion}}\quad we have demonstrated the capability of co-deceleration of lithium atom and \ce{CH3} molecule using Monte-Carlo simulation of the deceleration process in the moving trap decelerator, and have characterized their deceleration by comparing their deceleration efficiencies, revealing the dependence of the deceleration efficiency on the deceleration value. Our scheme offer several advantages over previous experiments: larger density of decelerated \ce{CH3} molecules, and co-trapping of title molecule and atom providing the possibility of study the collision properties of between them, thus open the door for investigating the prospects of sympathetic cooling. Many polyatomic free radicals in the doublet state have similar linear Zeeman effect as \ce{CH3} molecule since their spin-rotation interaction is typically smaller than the rotational spacing, thus can be Zeeman decelerated in the same way as we propose here for \ce{CH3}.

With our ongoing collisional study between lithium atoms and \ce{CH3} molecules, many promising applications will be enabled. For example, after loading them into a magnetic trap, the possibility of creating ultracold \ce{CH3} molecules by sympathetic cooling with ultracold lithium atoms can be stringently tested if elastic collision cross section, inelastic collision cross section, and reactive cross section between them are measured. Bimolecular collisions can also be studied inside such a trap, which has been shown for oxygen molecules \cite{stuhl2012evaporative,segev2019collisions}, opening an new avenue to investigate the possibility of evaporative cooling. Besides, study of cold reactions between excited lithium atoms and \ce{CH3} molecules are also possible according to previous theoretical calculations \cite{bililign2002nonradiative,hattaway2004energy}. With the ability of continuously changing the collision energy by tuning the trap depth, we can measure the reaction kinetics between them at the very low temperatures.

Yang Liu and Le Luo acknowledge helpful suggestion and discussion from Jiaming Li. Yang Liu acknowledge the financial support from National Natural Science Foundation of China(NSFC) under Grant No. 11974434, Fundamental Research Funds for the Central Universities of Education of China under Grant No. 191gpy276, Natural Science Foundation of Guangdong Province under Grant 2020A1515011159. Le Luo received supports from NSFC under Grant No.11774436, Guangdong Province Youth Talent Program under Grant No.2017GC010656, Sun Yat-sen University Core Technology Development Fund, and the Key-Area Research and Development Program of GuangDong Province under Grant No.2019B030330001.

\bibliography{MovingTrap_v4}

\end{document}